\documentclass[smallextended]{svjour3}

\smartqed
\usepackage{graphicx}
\usepackage{xcolor,dsfont}
\usepackage{appendix}
\usepackage{braket}
\usepackage{amsmath,bm}
\usepackage{revsymb4-1}
\usepackage{latexsym}
\usepackage{blindtext}
\usepackage{qcircuit}

\usepackage[colorinlistoftodos, color=green!40, prependcaption]{todonotes}
\usepackage{xcolor}
\usepackage{adjustbox}
\usepackage{placeins}
\usepackage[T1]{fontenc}
\usepackage{lipsum}
\usepackage{csquotes}
\usepackage{float}
\usepackage[colorlinks, breaklinks]{hyperref}
\usepackage{lineno}
\newsavebox{\definitionbox}
{\end{minipage}\end{lrbox}%
\begin{center}{\colorbox{checkcolor}{\usebox{\definitionbox}}}%
\end{center}}
\title{A complete quantum process model for physically realistic photo-detection in the
joint photon-number and spectral degrees of freedom}
\author{Sadia Rafi$^\dag$\and Aeysha Khalique $^{\dag\ddag\mathparagraph}$}
\institute{$\dag$ \at School of Natural Sciences, National University of Sciences and Technology,
	H-12 Islamabad, Pakistan
    \and $\ddag$ \at Department of Physics, Syed Babar Ali School of Science and Engineering, Lahore University of Management Sciences, Lahore, 54792, Pakistan
	\and $\mathparagraph$ \at National~Centre~for~Physics~(NCP),~Shahdra~Valley~Road,~Islamabad~44000,~Pakistan.} 
\date{Received: date / Accepted: date}
\date{\today}
\begin{document}

\maketitle
\begin{abstract}
Photo-detection is a fundamental component of experimental photonics, yet theoretical models often treat detectors solely in the photon-number basis, neglecting other relevant degrees of freedom such as spectral characteristics. We present a complete quantum process model for physically realistic photodetection that incorporates both photon-number effects and frequency-dependent detector response, described through explicitly derived Kraus operators in the operator-sum representation. This formalism accommodates realistic inefficiencies, models detection within finite spectral windows, and can be directly substituted for idealized models in quantum optical analyses. The model leads to identifying the suitable detection window of the detectors relative to the spectral distribution of incident photons and conversely, in order to maximize the observability of quantum effects.
\end{abstract}

\section{Introduction}
\label{sec:intro}
Photo-detectors are a vital part of quantum optics experiments and hence that of optics-based quantum information~\cite{BB84,BBC+93,PBW+98} and computation~\cite{KLM01,KMN+07}. Most of these processes rely on post-measurement results. Imperfect detectors limit the performance of these protocols~\cite{BLMS00,MQZ+05}. 
Although theoretical models often treat detectors purely in the photon-number basis, real photons possess additional degrees of freedom-most notably spectral/temporal modes-that significantly influence protocols. Realistic states, such as those from spontaneous parametric down-conversion or four-wave mixing, exhibit rich spectral-temporal correlations that affect interference, entanglement, and evolution through dispersive media ~\cite{rohde2005frequency,brecht2015photon,raymer2020temporal}. Yet, detection has often been modeled without fully accounting for these effects, despite their importance in phenomena like mode mismatch~\cite{rohde2005quantum,rohde2005optimal,Rohde2006}. Recent theoretical efforts have sought to generalize the standard POVM and beam-splitter models to account for photodetection in broadband or multimode settings~\cite{van2017time,Thomas2021,TualleBrouri2009}
. 

Modern detectors, including superconducting nanowire single-photon detectors (SNSPDs), transition-edge sensors (TES), and hybrid photodiodes, offer photon-number resolution with wavelength-dependent efficiencies~\cite{allevi2010photon,marsili2013detecting,lita2008counting}, and experiments show detection probabilities can vary significantly across the spectral bandwidth~\cite{kuzucu2008time,baragiola2012n}. Although quantum detector tomography~\cite{lundeen2009tomography} enables POVM reconstruction for photon number resolving (PNR) detectors, spectral dependencies are frequently neglected. A unified quantum process model that jointly treats photon-number and spectral domains is therefore essential.

Here, we present a physically motivated model that incorporates the spectral properties of both incident states and the detector’s frequency dependent response. Unlike idealized approaches assuming uniform detection, our formalism accounts for realistic inefficiencies and detection within finite spectral windows. Using the Kraus operator representation, we describe photodetection as positive operator valued measurement (POVM), enabling modular integration into quantum optical analyses. Our simplified results make this model to be used as a plug and play system in any photo detection process with initial states having photon number and spectral degrees of freedom.

We demonstrate the applicability of the model through three scenarios: single photon detection, multimode coherent states, and Hong–Ou–Mandel (HOM) interference, illustrating how it captures the interplay between photon statistics and spectral resolution under realistic conditions. Our observation that coherent states give the same detection effect as single photons if the spectral width of the incident beam is very large, suggests that such states can replace single photons in many realistic setups. The  observation of quantum effects, like HOM interference, are highly dependent on the spectral profile of detectors and the incident photons, specially the case when the photons become distinguishable based on their spectral profiles.

The paper is organized as follows. In Sec.~\ref{sec:Background}, we provide the details of existing photodetection model, in the context of joint spectral and photon number degrees of freedom, that forms the basis for our inefficient operator-based model. Sec.~\ref{sec:quaiKraus} introduces our proposed framework, where we construct a generalized model of photodetection based on the Kraus operator formalism. We derive the operator-sum representation of the detection process and explicitly develop its plug and play form for spectrally sensitive, inefficient multimode detectors. In Sec.~\ref{sec:examples}, we apply our model to a set of representative quantum optical scenarios to illustrate its versatility and physical relevance. These include: (i) the detection of a single photon with spectral selectivity, (ii) the spectral response of the detector to truncated coherent states, and (iii) the modeling of Hong-Ou-Mandel (HOM) interference with frequency dependent detection. Finally, we conclude in Sec.~\ref{sec:conclusion}.
\section{Background}\label{sec:Background}
The modeling of photodetection in quantum optics has evolved significantly, driven by the increasing precision of experiments and the practical limitations of physical detectors. This section traces the development of detector models from idealized descriptions to sophisticated quantum operations that incorporate spectral and photon-number degrees of freedom.
\subsection{Idealized photodetection model}
The ideal photo-detector is a device which responds to the photon-number of an incident state, and is able
to distinguish between different number states. Such a
detector can be modeled by the measurement projectors~\cite{helstrom1969quantum} 
\begin{align}
\Pi_n=\ket n\bra n ,
\end{align}
where $n$ is the measurement result, corresponding to
the number of photons in the incident state. Clearly
this set of projectors forms a legitimate POVM, since $\sum_{i=0}^{\infty}\Pi_n=\mathds1$.
 In general, detectors have flat spectral response. The single photon detection operator for such a detector is~\cite{Rohd2006}
\begin{align}
\Pi_1=\int \text d \omega a^\dagger(\omega)\ket 0\bra 0 a(\omega),
\end{align}
which generalises to the $n$-photon detection operator,
\begin{align}
\Pi_n = \frac{1}{n!} \int \text d^n\vec\omega a^{\dagger n}(\vec\omega)\ket { 0}\bra { 0 }a^n(\vec\omega),
 \end{align}
where $\text d^n\vec\omega=\text d\omega_1\text d\omega_2\cdots\text d\omega_n$,  $a^n(\vec\omega)=a(\omega_1)\cdots a(\omega_n)$ and $a^{\dagger n}(\vec\omega)=a^\dagger(\omega_1)\cdots a^\dagger(\omega_n)$.
\subsection{Realistic Detector Models}
\begin{figure}[!htbp]
\centering
	\includegraphics[width=0.4\linewidth]{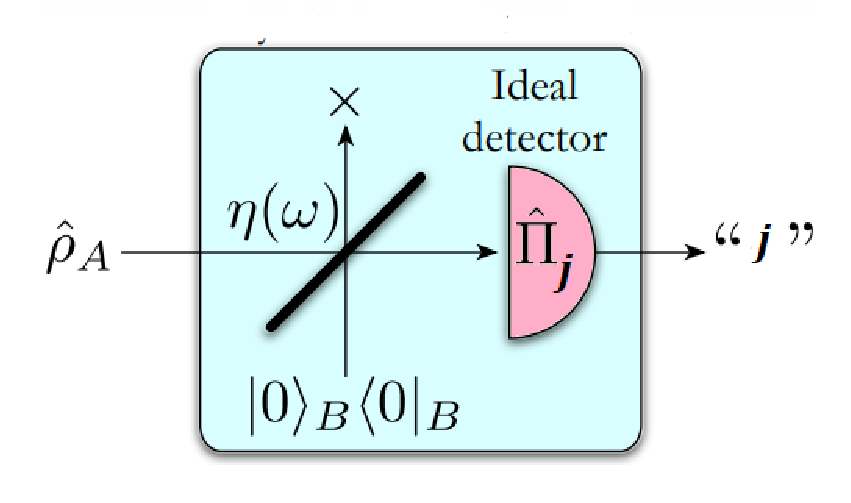}
\caption{
	An inefficient detector is modelled as a perfect detector preceded by a beam splitter with transmittivity  $\eta(\omega)$. The input $\rho_A$ is incident at one port and vacuum at the other port, which is discarded.
	 }
\label{fig:detector}
\end{figure}
An inefficient detector with efficiency $\eta(\omega)$ is modelled  by a perfect detector preceded by a beam splitter as in Fig.~\ref{fig:detector}, such that signal is incident on one port of the beam splitter and vacuum on the other port~\cite{Rohd2006}. The beam-splitter transformation matrix
\begin{align}
U_{AB}=
\begin{pmatrix}
 \sqrt{\eta(\omega) } & \sqrt{1-\eta(\omega) } \\
 \sqrt{1-\eta(\omega) } & -\sqrt{\eta(\omega) }
 \end{pmatrix}
 \label{eq:UAB}
\end{align}
  transforms the input photon creation operators as 
{\small
$a^\dagger(\omega)\xrightarrow{U} \sqrt{\eta(\omega) }a^\dagger(\omega)+ \sqrt{1-\eta(\omega) } b^\dagger(\omega)$} where $b$ is the mode at the other port and it transforms as
 {\small
 $b^\dagger(\omega)\xrightarrow{U} \sqrt{1-\eta(\omega) }a^\dagger(\omega)- \sqrt{\eta(\omega) } b^\dagger(\omega)$.} 
We refine this model to include the spectrally resolved detector for detection of any input state $\rho_A$ and the transformation of the remaining quantum state.
\section{Kraus operator for photo detection}
\label{sec:quaiKraus}
Photo detection model for spectrally resolved states takes into account the spectral profile of the incident state and the detector. We present the detection operation based on the beam splitter model presented in Sec.~\ref{sec:Background}.  In this section, we present our model of spectrally resolved inefficient photodetection and give a quantum map of the process in terms of Kraus operators. 

The detection of $j$ photons by imperfect photodetector is modeled as in Fig.~1. Beam splitter acts on combined state of $A$ and $B$. The reflected mode $B$ is traced out followed by perfect detection of mode $A$ by projection operator $\Pi_{j,A}$ detecting $j$ photons. The unnormalized state $\Phi_j$ after detection is 
\begin{align}
\Phi_j&=\Pi_{j,A}\text{Tr}_B\left\{U_{AB}(\rho_A\otimes\ket{0}\bra{0}_B) U_{AB}^\dagger \right\}\Pi_{j,A}\nonumber\\
&=\text{Tr}_B\left\{U_{AB}U^\dagger_{AB}\Pi_{j,A}U_{AB}(\rho_A\otimes\ket{0}\bra{0}_B) U_{AB}^\dagger \Pi_{j,A}U_{AB}U^\dagger_{AB}\right\}
\label{eq:detectormap}
\end{align}
The normalization $\rm {Tr}(\Phi_j)$ gives the probability of detecting $j$ photons.
 The transformed projection operator \begin{align}\Pi'_{j,AB}=U^\dagger_{AB}\left(\Pi_{j,A}\otimes I_B\right) U_{AB}
     \label{Eq:PijPrime}
 \end{align}
and the trace on the reflected mode $B$, leads to the unnormalized state
\begin{align}
\Phi_j=\sum_{N}\int\text d^N\vec\omega K_N(\vec\omega,j )\rho_A K_N^\dagger(\vec\omega,j).
\label{eq:map}
\end{align}
with $\text d^N(\vec\omega)=\text  d\omega_1\cdots\text d\omega_N$ and
\begin{align}
K_N(\vec\omega,j)=\frac{1}{\sqrt{N!}}{}_B\bra 0 b^N(\vec\omega)U_{AB}\Pi'_{j,AB}\ket 0{}_B.
\label{eq:KNGen}
\end{align}
Here, $b^N(\vec\omega):=b(\omega_1)\cdots b(\omega_N)$ and $\{b(\omega_i);i=1,\dots, N\}$ are annihilation operators of the reflected mode $B$.  The Kraus operators $K_N(\vec\omega,j)$ are only a subset of operators spanning a subspace for particular $j$. For a complete set of operators 
$\sum_j\sum_NK^\dagger_N(\vec\omega,j)K_N(\vec\omega,j)=\mathds1$.

The explicit calculation of the Kraus operators from Eq.~(\ref{eq:KNGen}) is given in Appendix~\ref{sec:appendixA}, leading to Kraus operators $K_N(\vec\omega,j)$ in Eq.~(\ref{eq:KNGenApp}). For $i$ incident photons on the detector, the sum of $j$ detected and $m$ missed photons must be equal to $i$ incident photons, leading to simplified modified Kraus operators in Eq.~(\ref{eq:ModKNiApp})
\begin{align}
    K_N(\vec\omega,j,i)
    &=P_N(\vec\omega)\frac{\delta_{N,i-j}}{j!N!\sqrt{N!}}\int \text d^j(\vec\Omega)a^{\dagger j}(\vec\Omega)\ket 0\bra 0 a^j_{\sqrt{\eta}}(\vec\Omega)a^N_{\sqrt{1-\eta}}(\vec{\omega}).
    \label{eq:KNi}
\end{align}   
given here again for completeness. The annihilation operators $a^{j}_{\sqrt{\eta}}(\vec{\Omega})$ stand for $j$ factors of the type $\sqrt{\eta(\Omega_r})a(\Omega_r)$ with $r\in \{1,\cdots,j\}$. A similar explanation is true for $a^N_{\sqrt{1-\eta}}(\vec{\omega})$. $P_N(\vec\omega)$ is $N$ permutations of $\omega_i$ for $i\in \{1,\ldots,N\}$.

Above Kraus operators, $K_N(\vec\omega,j,i)$, take into account the detector inefficiencies and the spectral profile of the detector. This is a plug and play model and can be used in any setup with incident photons having photon number and spectral degrees of freedom.
\section{Applications of multimode detector}
\label{sec:examples}
We have developed the Kraus operator representation of photon detection. This model of an inefficient realistic photon detector can detect any incident state $\rho_A$ with photon number and spectral degrees of freedom. We apply the model to the detection of single photon and multiphoton states and find the appropriate spectral profile of the incident state for a detector and vice a versa. We consider three examples: detection of a single photon state in Sec.~\ref{sec:example_sing}, coherent state in Sec.~\ref{sec:COH} and Hong-Ou-Mandel dip in Sec.~\ref{sec:HOM}.
\subsection{Detection of single photon state}
\label{sec:example_sing}
A single photon pulse, with $i=1$, having spectral profile $\psi(\beta)$ and state
\begin{align}
    \rho_A=\iint\text d\beta\text d\beta'\psi(\beta)a^\dagger(\beta)\ket 0_A\bra 0 a(\beta') \psi^*(\beta')
\end{align}
is incident on the detector with transmitivity $\eta(\Omega)$. From Eq.~(\ref{eq:KNi}), for $i=1$, detection of $j=1$  photon corresponds to the modified multimode Kraus operators
\begin{align}
    K_0(\omega,1,1)&=\int\text d\Omega \sqrt{\eta(\Omega)} a^\dagger (\Omega)\ket 0_A\bra 0 a(\Omega),\nonumber\\
    K_1(\omega,1,1)&=0,
\end{align}
lead to the state
\begin{align}
\Phi_j(\rho_A)=A^\dagger_{|\eta|\psi}\ket 0\bra 0 A_{|\eta|\psi*},
\end{align}
where 
\begin{align}
    A^\dagger_{|\eta|\psi}\equiv\int \text d\Omega\sqrt{\eta(\Omega)}\psi(\Omega)a^\dagger(\Omega).
    \label{eq:A}
\end{align}
The probability of detecting $j=1$ photon is
\begin{align}
P_{j=1}=\int \text d\Omega\eta(\Omega)\lvert{\psi(\Omega)}\rvert^2.
\label{eq:Pj1}
\end{align}
The probability of detection of the photon depends on the overlap of the transmitivity $\eta(\Omega)$ of the detector and the spectral profile $\psi(\Omega)$ of the incident photon.

The probability of detection of $j=0$ photon is computed from the modified Kraus operators $K_N(\omega,j,1)$ in Eq.~(\ref{eq:KNi}). For $N=0$ and $N=1$,
\begin{align}
    K_0(\omega,0,1)&=0,\\
    K_1(\omega,0,1)&= \sqrt{1-\eta(\omega)}\ket 0_A\bra 0 a(\omega),\nonumber
    \label{eq:K0K1}
\end{align}
lead to the post detection state
\begin{align}
\Phi_j(\rho)=\int\text  d\omega\sqrt{1-\eta(\omega)}\psi(\omega)\ket 0_A\bra 0 \sqrt{1-\eta(\omega)}\psi^*(\omega).
\end{align}
The probability of $j=0$ photons is thus
\begin{align}
P_{j=0}=\int \text d\omega(1-\eta(\omega))\lvert{\psi(\omega)}\rvert^2.
\label{eq:Pj0}
\end{align}

\begin{figure}[!htbp]
    \centering    \includegraphics[width=0.45\linewidth]{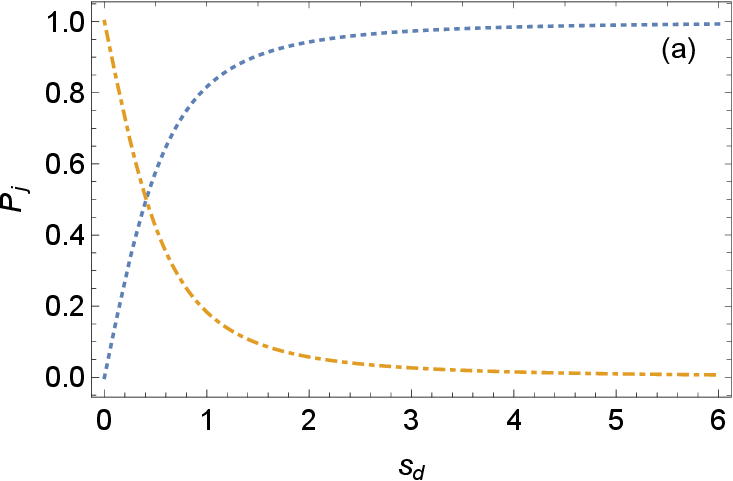}\;\;\;\includegraphics[width=0.45\linewidth]{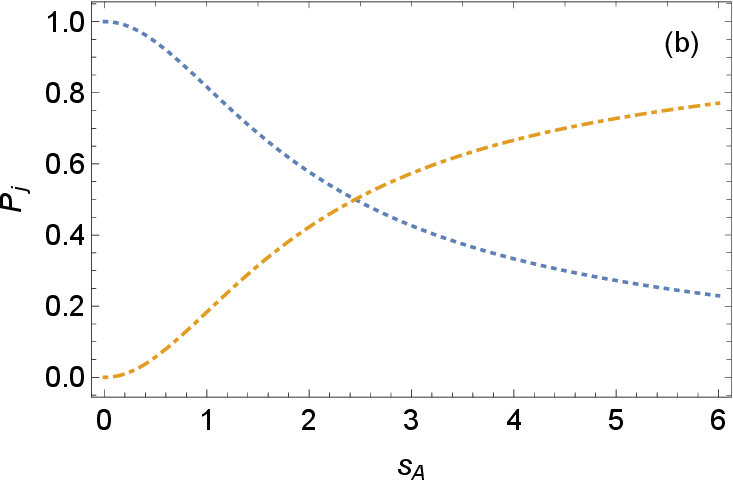}
    \caption{Probability $P_j$ of detection of $j=1$ photon (dotted curve) and $j=0$ photon (dot-dashed curve) by (a) varying detector spectral bandwidth $s_d$ for fixed photon spectral bandwidth $s_A=1$ (b) varying photon spectral bandwidth $s_A$ for fixed detector spectral bandwidth $s_d=1$. Here $\mu=0$}  
\label{fig:Single}
\end{figure}
The probability $P_{j}$ of detection of the incident photon from Eq.~(\ref{eq:Pj1}) and that of detection of no photon from Eq.~(\ref{eq:Pj0}) is plotted in Fig.~\ref{fig:Single}. Here, the incident single photon pulse and detector spectral profile are Gaussian with spectral bandwidths $s_A$ and $s_d$, respectively
\begin{align}
    \psi(\omega)=\frac{1}{(2  \pi s_A^2)^{1/4}}\text e^{-(\omega-\mu)^2/(4 s_A^2)},\;\;\;\eta(\omega)=\text e^{-\omega^2/(4 s_d^2)}.
    \label{eq:psi}
\end{align} 
In Fig.~\ref{fig:Single}(a), photon spectral bandwidth is fixed at $s_A=1$, and detector spectral bandwidth $s_d$ is varied. Initially, at $s_d=0$ no photon is detected. As $s_d$ increases, the photon detection probability increases and reaches 1 as soon as the detector's spectral bandwidth is more than that of the incident photon $s_A=1$. The increase is slow if the spectral bandwidth of photon $s_A$ is large. In Fig.~\ref{fig:Single}(b) $s_d=1$ and $s_A$ is varied. Initially, for $s_A$ much smaller than $s_d$, the photon is detected with 100\% probability $P_{j=1}=1$. As the photon spectral bandwidth increases, it becomes less probable to be detected and $P_{j=0}$ increases. However, the decrease in detecting $j=1$ is much slower than the decrease of $j=0$ in Fig.~\ref{fig:Single}(a). This implies that a photon having spectral profile aligned with the detector, with $\mu=0$, has a certain probability of detection, even if the spectral bandwidth of the photon is much larger than that of the detector.

\subsection{Detector Response for Coherent States}
\label{sec:COH}
We apply our developed model to compute photon detection probabilities \( P_{j} \) for an input coherent state.

Consider a multimode, Poisson distributed coherent state \cite{glauber1963,loudon2000quantum}

\begin{align}
\rho_{A}=\ket{\alpha}\bra{\alpha}= \text e^{\int\text d\xi\lvert{\alpha{(\xi)}}\rvert^2}
\sum_{i=0}^{\infty}\sum_{i'=0}^{\infty}\int\int \text d^i\vec\beta \text d^{i'}\vec\beta' \frac{1}{i!i'!}a_{\alpha}^{\dagger i}(\vec\beta)\ket{0}\bra{0} a_{\alpha}^{i'}(\vec\beta'),
\label{eq:coherentRho}
\end{align}
 where $a^i_\alpha(\vec\beta)$ is $i$ factors of $\alpha(\beta_r)a^\dagger(\beta_r)$ for $r\in\{1,i\}$, where $\alpha(\beta_r)$ is the spectral profile of the photons. The mean photon number of the coherent state is $\langle \hat{n} \rangle=\int\text d\xi\lvert{\alpha{(\xi)}}\rvert^2$
 
The modified Kraus operators from Eq.~(\ref{eq:KNi}) acting on the $i$th term in $\rho_A$ give the probability of detecting $j$ photons,
\begin{align}
 P_{j} = \text e^{-\int\text d\xi\lvert{\alpha{(\xi)}}\rvert^2}\sum_{i=j}^{\infty} \frac{1}{j! (i - j)! }I_{1}^{j}I_{2}^{i-j},
\end{align} which depends on $I_1= \int\text d\omega\lvert{\alpha(\omega)}\rvert^2 \eta(\omega)$ and $I_2= \int\text d\omega\lvert{\alpha(\omega)}\rvert^2 (1 - \eta(\omega))$. We see that $I_2=1-I_1$.

\begin{figure}[!htbp]
 	\centering
 	\includegraphics[width=0.45\linewidth]{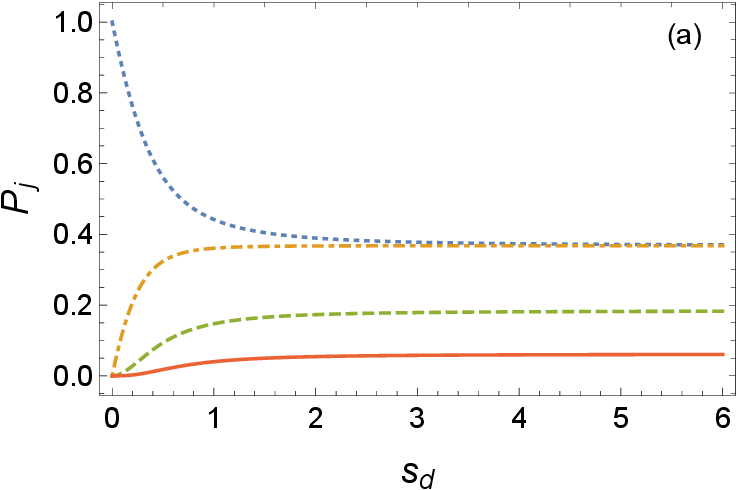}
    \;\;\;\includegraphics[width=0.45\linewidth]{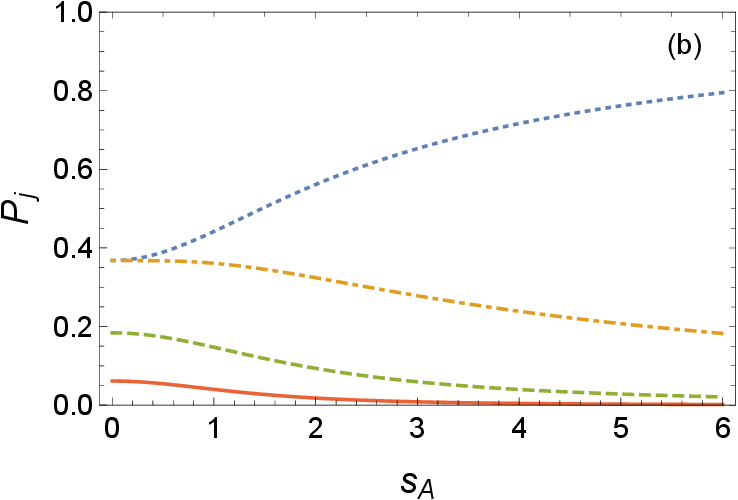}
 	\caption{Probability $P_j$ of detection of $j=0$ photon (dotted curve), $j=1$ photon (dot-dashed curve), $j=2$ photons(dashed) and $j=3$ (solid curve) in (a) by varying detector spectral bandwidth $s_d$, for photon spectral bandwidth $s_A=1$, and in (b) by varying photon spectral bandwidth $s_A$ for $s_d=1$. $\mu=0$ for both.}
    \label{fig:cohprob}
 \end{figure}

The probability $P_j$ against detector spectral bandwidth $s_d$ is plotted in Fig.~\ref{fig:cohprob}(a) for $j=0,1,2$ and 3 photons. At $s_d=0$ the detector is "infinitely selective" and cannot capture any photons from the coherent state.
As the detection window widens by increasing $s_d$, the main contribution in detection probability comes from $j=0$ and $j=1$ photon, due to the Poisson-distributed photon field (coherent state), which approach the same value asymptotically. The probability of detection of $j=2$ and more photons, from the exponential tail of the coherent state, also approaches a fixed small value asymptotically. This implies that for coherent state to detect a photon, increasing the spectral width of the detector does not guarantee the detection of a photon. Fixing $s_d=1$ and increasing the spectral bandwidth of the coherent state, in Fig.~\ref{fig:cohprob}(b) further decreases the probability of detecting a photon or more. 

Comparing Fig.~\ref{fig:Single}(b) and \ref{fig:cohprob}(b), we see that the detection probabilities for coherent state approaches that of a single photon with the increase of spectral bandwidth of the incident photons. This implies that if $s_A$ is very large, we can use coherent state with the same detection probabilities as that of a single photon, where coherent states are much easier to generate experimentally than single photons.
\subsection{Hong-Ou-Mandel Dip}
\label{sec:HOM}
\begin{figure}[!htbp]
 	\centering 	\includegraphics[width=0.4\linewidth]{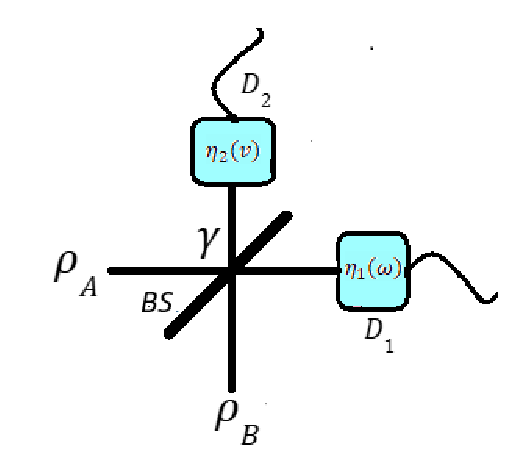}
 	\caption{Hong-Ou-Mandel effect}
    \label{fig:HOM}
 \end{figure}

The theoretical framework applied in this section is grounded in the foundational understanding of two-photon interference. Two incident photons with states $\rho_A$ and $\rho_B$ are incident on a beam splitter with transmitivity $\gamma$, and the probability of detection of photons is measured at the two detectors, $D_1$ and $D_2$ as shown in Fig~\ref{fig:HOM}. The dip in coincidence probability on the two detectors for indistinguishable photons~\cite{hong1987measurement} is an interplay of quantum interference. 
 
We consider initial states to be single photons with a frequency profile $\psi(\beta)$ and $\zeta(\delta)$ 
\begin{align}
    \rho_A &=\iint\text d\delta \text d\delta' \text e^{\text iT \delta}\psi(\delta)a^\dagger(\delta)\ket 0\bra 0 a(\delta')\text e^{-\text i T \delta'}\psi^*(\delta')\nonumber\\
\rho_B&=\iint\text d\beta \text d\beta' \zeta(\beta)b^\dagger(\beta)\ket 0\bra 0 b(\beta')\zeta^*(\beta')
    \end{align}
with photon $A$ having a time delay $T$. The state $\rho'$ at the two output ports of the beam splitter with transmitivity $\lambda$ contains the combination of both photons at one port and one at each
\begin{align}
    \rho'=&\iint\text d\beta \text d\delta\text e^{- \text i T\delta}\zeta(\beta)\psi(\delta)\Big[\sqrt{\gamma(1-\gamma)}a^\dagger(\beta)a^\dagger({\delta})+(1-\gamma) a^\dagger(\delta)b^\dagger(\beta)
    -\gamma b^\dagger(\delta)a^\dagger(\beta)\nonumber\\&-\sqrt{\gamma(1-\gamma)} b^\dagger(\beta)b^\dagger({\delta})\Big]
    \ket{00}\bra{00}\iint\text d\beta' \text d\delta'\text e^{\text iT\delta'}\zeta(\beta')\psi(\delta')\Big[\sqrt{\gamma(1-\gamma)}a(\beta')a({\delta'})\nonumber\\&+(1-\gamma) a(\delta')b(\beta')
    -\gamma b(\delta')a(\beta')-\sqrt{\gamma(1-\gamma)}b(\beta')b({\delta'})\Big]
    \label{eq:rhoPHOM}
\end{align}

For above two-photon input state with $i=2$, the conditional output state corresponding to the detection of $j$ photons at detector $D_1$ and $j'$ photons at detector $D_2$ is given by
\begin{align}
\Phi_{jj'}=\sum_{M=0}^2\sum_{N=0}^2\iint \text d^N(\vec\omega)\text d^M(\vec v)
K_N(\vec \omega,j,2)
K_M(\vec v,j',2)
\rho'
K^\dagger_N(\vec \omega,j,2)
K^\dagger_M(\vec v,j',2).
\label{eq:phijj}
\end{align}
Here, $K_N(\vec\omega,j,2)$ denotes the modified Kraus operator associated with detector $D_1$, as defined in Eq.~(\ref{eq:KNi}), for a detector characterized by the frequency-dependent detection efficiency $\eta_1(\vec\omega)$. Similarly, $K_M(\vec v,j',2)$ describes detector $D_2$ and is obtained from $K_N(\vec\omega,j,2)$ by replacing the mode operators $a$ and $a^\dagger$ with $b$ and $b^\dagger$, respectively, with detection efficiency $\eta_2(\vec v)$. Eq.~(\ref{eq:phijj}) therefore represents the unnormalized conditional state of the system following the joint measurement outcome $jj'$ at the two detectors. The  detection probabilities for detection of $j$ photons on detector $D_1$ and $j'$ photons on detector $D_2$, 
\begin{equation}
P_{jj'}={\rm Tr}({\Phi_{jj'}})
\label{eq:probhom}
\end{equation}
are analyzed for various spectral profiles of incident states and the detectors.

\subsubsection{Discussion}
We find the detection probabilities $P_{jj'}$ using~Eqs.~(\ref{eq:rhoPHOM})~-~(\ref{eq:probhom}). Here $\eta_{i}(\omega)$ and $\psi(\delta)=\zeta(\delta)$ are Gaussian as in Eq.~(\ref{eq:psi}) and $s_{d_{i}}$ and $s_{A_{i}}$ for $i=1,2$ are the spectral bandwidths of the two detectors and frequency profiles for incident photons, respectively. Possible detection probabilities are $P_{20}$ (both photons detected on detector $D_1$), $P_{02}$ (both photons detected on detector $D_2$), $P_{11}$ (one photon detected on each detector), $P_{10}$ (one photon detected on detector $D_1$ and none on the other), $P_{01}$ (one photon detected on detector $D_2$ and none on the other) and $P_{00}$ (no photon detected on either detector). In each case, the sum of all the probabilities, $\sum P_{jj'}=1$.
\begin{figure}[!htbp]
    \centering
\includegraphics[width=0.45\linewidth]{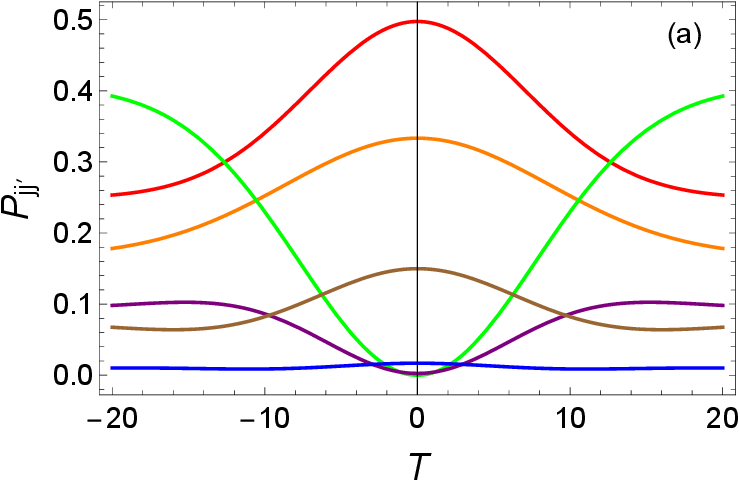}\;\;\;\includegraphics[width=0.45\linewidth]{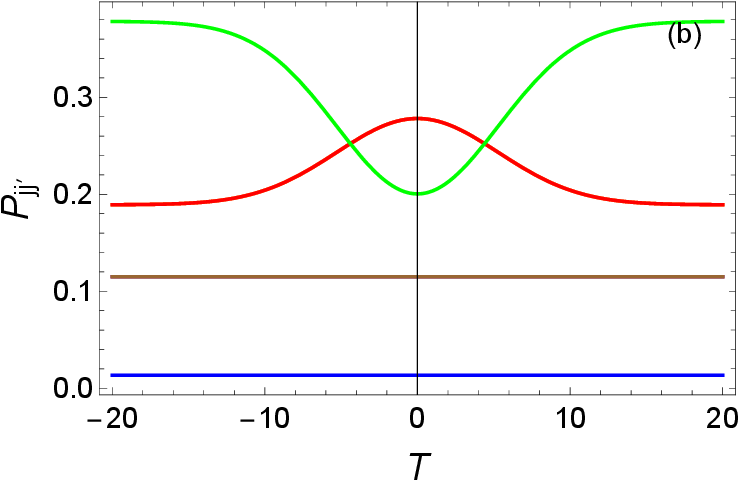}
    \caption{Probabilities $P_{jj'}$ of detection of $j$ photons on one detector and $j'$ on the other vs the time delay $T$ for $P_{20}$ (orange), $P_{02}$ (red), $P_{11}$ (green), $P_{01}$ (purple), $P_{10}$ (brown) and $P_{00}$ (blue) curve for\\
    (a) $s_{d_1}=0.1$,  $s_{d_2}=1.0$, $s_{A_1}=s_{A_2}=0.1$ and $\mu_1=\mu_2=0$ and\\ (b) $s_{d_1}=s_{d_2}=0.4$, $s_{A_1}=0.1$, $s_{A_2}=0.4$,  $\mu_1=0.2$ and $\mu_2=0$.} 
    \label{fig:Pjj}
\end{figure}

In Fig.~\ref{fig:Pjj}(a), we plot the probabilities $P_{jj'}$ vs the time dealy $T$. The two photons are identical otherwise with $s_{A_1}=s_{A_2}$ and $\mu_1=\mu_2$. For the detectors having different bandwidths with $s_{d_2}=10\times s_{d_1}\gg s_{A_i}$, the bunching probability $P_{02}=0.5$ (red curve) at zero time delay and attains a constant value of 0.25, as is the case of single mode two photon interference. $P_{20}<0.5$ (orange curve) at zero time delay, and attains a constant value $<0.25$, as the detector $D_1$ is selective. We see a dip in coincidence probability $P_{11}$ (green curve) at $T=0$, which is the characteristic Hong-Ou-Mandel (HOM) dip. It attains a constant value of $0.4$ as the time delay increases, which is less than $0.5$ for single mode two photon interference, as there is significant contribution from $P_{01}$ (purple) and $P_{10}$ (brown curve). $P_{00}$ (blue curve) stays close to zero.

In Fig.~\ref{fig:Pjj}(b) the two incident photons are distinguishable with different spectral bandwidths, $s_{A_2}=4\times s_{A_1}$, and means, $\mu_1=0.2$ and $\mu_2=0$, with the same color coding as Fig.~\ref{fig:Pjj}(a). The two detectors are identical with bandwidth comparable to $s_{A_2}$. The identical detectors lead to $P_{20}=P_{02}$ (red curve) and $P_{01}=P_{10}$. The distinguishability of the photons based on their spectral profiles, however, results in a shallow dip in coincidence, $P_{11}$, at zero time delay. The bunching probabilities $P_{20}=P_{02}$ are also much less than the peak value of $0.5$ at $T=0$. The HOM dip is thus highly dependent on the spectral degree of freedom of the incident photons. For very high distinguishability $s_{A_1}\ll s_{A_2}$, $P_{11}>P_{20}$ at all $T$, although both anti-bunching and bunching probabilities still show a shallow dip and peak, respectively. However, $P_{00}$ stays close to zero, as the spectral bandwidth of the detectors in this case is wide enough to detect at least one photon. $P_{00}$ becomes the most dominating probability if $s_{d_i}\ll s_{A_j}$ for $i,j\in{1,2}$.

The HOM interference showcases the interplay between photon statistics and spectral resolution. This leads to accurate predictions about the experimental outcomes.
\section{Conclusion}
\label{sec:conclusion}
We introduce a simplified, physically motivated, comprehensive model for photodetection that accounts for factors such as photon number and spectral degree of freedom. This operator sum representation offers a practical, interchangeable model for realistic photo detection, serving as a replacement for more simplistic existing models. It can be used as a plug and play model in any detection process involving incident photons with photon number and spectral degrees of freedom. This model is versatile and can be applied to various photonic degrees of freedom, including spectral/temporal, spatial, and polarization aspects. 

To demonstrate the utility of our detection model, we apply it to three canonical quantum optical scenarios: single-photon detection, detection of a truncated multimode coherent state, and the Hong-Ou-Mandel (HOM) interference experiment. In each case, the model captures the interaction between photon-number statistics and spectral resolution under realistic conditions. It gives the suitable detection window of the detector relative to the spectral width of incident photons, to maximize the observation of quantum affects. In Hong-Ou-Mandel dip, the model captures the distinguishability of the incident photons based on their spectral profile, leading to shallow dips. 

This model acts as a plug-and-play model for realistic photo-detection. Having such a model for photodetection, is essential for advancing modern and realistic quantum optics and optical quantum information processing.

\appendix
\section*{Appendix}
\setcounter{section}{0}
\renewcommand{\thesection}{\Alph{section}}

\numberwithin{equation}{section}

\section{Explicit Calculation of Kraus Operators} 
\label{sec:appendixA}
We give the explicit calculation of Kraus operators from Eq.~(\ref{eq:KNGen}) 
\begin{align}
K_N(\vec\omega,j)=\frac{1}{\sqrt{N!}}{}_B\bra 0 b^N(\vec\omega)U_{AB}\Pi'_{j,AB}\ket 0{}_B,
\label{eq:KNGenApp}
\end{align}
with 
\begin{align}\Pi'_{j,AB}=U^\dagger_{AB}\left(\Pi_{j,A}\otimes I_B\right) U_{AB}
     \label{Eq:PijPrimeApp}
 \end{align}
 from Eq.~(\ref{Eq:PijPrime}).
 
The perfect projection operator on mode $A$
\begin{align}
\Pi_{j,A} = \frac{1}{j!} \int \text d^j\vec \Omega a^{\dagger j}(\vec\Omega)\ket 0\bra0a^j(\vec\Omega),
\label{eq:PijA}
\end{align}
where $\text d^j\vec\Omega=\text d\Omega_1\text d\Omega_2\cdots\text d\Omega_j$.
For mode $B$ on the fictitious beam splitter
\begin{align}
    I_B &=  \sum_{m=0}^\infty \frac{1}{m!} \int \text d^m\vec \gamma b^{\dagger m}(\vec\gamma)\ket 0\bra 0 b^m(\vec\gamma),
    \label{eq:PiB}
\end{align}
Using above equations and the beam splitter $U_{AB}$ from Eq.~(\ref{eq:UAB}), the transformed projection operator in Eq.~(\ref{Eq:PijPrimeApp})is
\begin{align}
&\Pi'_{j,AB}=U^\dagger_{AB}\left(\Pi_{j,A}\otimes I_B\right)U_{AB}\nonumber\\
&=\sum_{m=0}^\infty\frac{1}{j!m!}\int\text d^j\vec{\Omega}\sum_{k=0}^j\binom{j}{k}a^{\dagger k}_{\sqrt{\eta}}(\vec\Omega)b^{\dagger (j-k)}_{\sqrt{1-\eta}}(\vec\Omega) \int\text d^m(\vec \gamma)\sum_{n=0}^m\binom{m}{n}a^{\dagger n}_{\sqrt{1-\eta}}(\vec\gamma)b^{\dagger (m-n)}_{\sqrt{\eta}}(\vec\gamma)(-1)^{m-n}\nonumber\\
&\ket{00}_{AB}\bra{00}
\sum_{k''=0}^j\binom{j}{k''}a^{k''}_{\sqrt{\eta^*}}(\vec \Omega)b^{(j-k'')}_{\sqrt{1-\eta^*}}(\vec \Omega)\sum_{n''=0}^m\binom{m}{n''}a^{n''}_{\sqrt{1-\eta^*}}(\vec \gamma)b^{(m-n'')}_{\sqrt{\eta^*}}(\vec \gamma).
\label{eq:PijPrimefullApp}
\end{align}
Here $a^{\dagger k}_{\sqrt{\eta}}(\vec{\Omega})$ stands for $n$ factors of the type $\eta(\Omega_r)a^\dagger(\Omega_r)$ with $r\in \{1,\cdots,n\}$. Similar explanation is true for all other operators in above expression. Note that none of the indices occurring in these factors is ever repeated, and all indices appear exactly once.

Now $\Pi''_{j,AB}=U_{AB}\Pi'_{j,AB}$ is 
\begin{align}
  &\Pi''_{j,AB}\nonumber\\
  &=\sum_{m=0}^\infty\frac{1}{j!m!}\int\text d^j\vec{\Omega}\sum_{k=0}^j\binom{j}{k}\sum_{l=0}^k\binom{k}{l}a^{\dagger l}_{{\eta}}(\vec\Omega)b^{\dagger (k-l)}_{\sqrt{\eta(1-\eta)}}(\vec\Omega) \sum_{s=0}^{j-k}\binom{j-k}{s}a^{\dagger s}_{\sqrt{\eta(1-\eta)}}(\vec\Omega)b^{\dagger (j-k-s)}_{{1-\eta}}(\vec\Omega)\nonumber\\
  &\int\text d^m(\vec \gamma)\sum_{n=0}^m\binom{m}{n}\sum_{p=0}^n\binom{n}{p}a^{\dagger p}_{\sqrt{\eta(1-\eta)}}(\vec\gamma)b^{\dagger (n-p)}_{\sqrt{1-\eta)}}(\vec\gamma)\sum_{q=0}^n\binom{m-n}{q}a^{\dagger q}_{\sqrt{\eta(1-\eta)}}(\vec\gamma)b^{\dagger (m-n-q)}_{\eta}(\vec\gamma)\nonumber\\
  &(-1)^{j-k-s+m-n-q}\nonumber\\
&\ket{00}_{AB}\bra{00}
\sum_{k''=0}^j\binom{j}{k''}a^{k''}_{\sqrt{\eta^*}}(\vec \Omega)b^{(j-k'')}_{\sqrt{1-\eta^*}}(\vec \Omega)\sum_{n''=0}^m\binom{m}{n''}a^{n''}_{\sqrt{1-\eta^*}}(\vec \gamma)b^{(m-n'')}_{\sqrt{\eta^*}}(\vec \gamma)
\label{eq:Pijpp}
\end{align}
Inserting above in Eq.~(\ref{eq:KNGenApp}), the Kraus operators $K_N(\vec\omega,j)$ take the explicit form
\begin{align}
 K_N(\vec\omega,j)&=\frac{1}{j!\sqrt{N!}}\sum_{m}\frac{1}{m!}\sum_{k,l,s,n,p,q=0}^{j,k,j-k,m,n,m-n}
\delta_{N,j+m-l-s-p-q}CH(\eta,\vec\omega)\nonumber\\
&\times\iint\text d^{s+l}\vec{\Omega}\text d^{p+q}\vec\gamma A_1^{\dagger}(\eta,\vec\Omega,\vec\gamma)\ket 0_A\bra 0 A_2(\eta^*,\vec\Omega,\vec\omega,\vec\gamma)P_N(\vec\omega),
\label{eq:KNFullApp}
\end{align}
with
\begin{align}
    C&=\binom{j}{k}\binom{k}{l}\binom{j-k}{s}\binom{m}{n}\binom{n}{p}\binom{m-n}{q}(-1)^{j-k-s-q}\nonumber\\
    H(\eta,\vec\omega)&=\sqrt{\eta(\vec\omega)(1-\eta(\vec\omega))}^{(j-s-l)}
(\eta(\vec\omega))^{(m-n-q)}(1-\eta(\vec\omega))^{(n-p)}, \nonumber\\
A_1^{\dagger}(\eta,\vec\Omega,\vec\gamma)&=a^{\dagger l}_{{\eta}}(\vec\Omega)a^{\dagger s}_{{1-\eta}}(\vec\Omega)a^{\dagger p+q}_{\sqrt{\eta(1-\eta)}}(\vec \gamma),\nonumber\\
A_2(\eta^*,\vec\Omega,\vec\omega,\vec \gamma)&=a^{s+l}_{\sqrt{\eta^*}}(\vec{\Omega})a^{j-s-l}_{\sqrt{\eta^*}}(\vec{\omega})a^{m-p-q}_{\sqrt{1-\eta^*}}(\vec{\omega})a^{p+q}_{\sqrt{1-\eta^*}}(\vec{\gamma}),
\label{eq:KNfactorsApp}
\end{align}
and $P_N(\vec\omega)$ is $N$ permutations of $\omega_i$ for $i\in \{1,\ldots,N\}$. In $H(\eta,\vec\omega)$, $\sqrt{\eta(\vec\omega)(1-\eta(\vec\omega))}^{(j-s-l)}$ is a short hand for $j-s-l$ factors of the type $\sqrt{\eta(\omega_r)(1-\eta(\omega_r))}$, where $r\in\{1,\ldots, N\}$. Similar notations are used for other two factors in $H(\eta,\vec\omega)$.  $a^{\dagger l}_{\sqrt{\eta}}(\vec{\Omega})$  and all other operators $A_1^{\dagger}(\eta,\vec\Omega,\vec\gamma)$ and $A_2(\eta^*,\vec\Omega,\vec\omega,\vec \gamma)$ have similar explanation as in Eq.~(\ref{eq:PijPrimefullApp}). Note that none of the indices occurring in these factors is ever repeated, and all indices appear exactly once on the left and the right side of $\ket0_A\bra0$.     

In Kraus operator, when $A_2(\eta^*,\vec\Omega,\vec\omega,\vec \gamma)$ in Eq.~(\ref{eq:KNfactorsApp}) operates on an incident Fock basis $a^i(\beta)\ket 0$, it must satisfy $\delta_{i,j+m}$ leading to the modified Kraus operators
\begin{align}    K_N(\vec\omega,j,i)&=\delta_{i,j+m}K_N(\vec\omega,j).
\end{align}   
This condition in fact means that the sum of the detected and the missed photons is equal to the number of incident photons. The modified Kraus operators using Mathematica code, simply reduces to
\begin{align}    K_N(\vec\omega,j,i)
    &=P_N(\vec\omega)\frac{\delta_{N,i-j}}{j!N!\sqrt{N!}}\int \text d^j(\vec\Omega)a^{\dagger j}(\vec\Omega)\ket 0\bra 0 a^j_{\sqrt{\eta}}(\vec\Omega)a^N_{\sqrt{1-\eta}}(\vec{\omega}).
\label{eq:ModKNiApp}
\end{align}
These modified Kraus operators can be used as a plug and play model in any detection operation involving number states.
\begin{acknowledgements}
The authors thank Peter P. Rohde and B. C. Sanders for many fruitful discussions and helpful insights.
\end{acknowledgements}

\bibliographystyle{spphys}  
\bibliography{references} 
\end{document}